\documentclass[preprint,aps]{revtex4}
\usepackage[dvips]{graphicx}
\usepackage{amsmath}
\begin{document}

\preprint{preprint}

\title{Formation of spatial patterns in an epidemic model with constant removal rate of the infectives}% Force line breaks with \\
\author{Quan-Xing Liu}
%\email{liuqx315@sina.com}%Lines break automatically or can be forced with \\
\author{Zhen Jin}%
 %\email{jinzhn@263.net}

\affiliation{%
Department of mathematics, North University of China,\\
Taiyuan, Shan'xi, 030051, People's Republic of China
}%

\date{\today}% It is always \today, today,

\begin{abstract}
This paper addresses the question of how population diffusion
affects the formation of the spatial patterns in the spatial
epidemic model by Turing mechanisms. In particular, we present
theoretical analysis to results of the numerical simulations in two
dimensions. Moreover, there is a critical value for the system
within the linear regime. Below the critical value the spatial
patterns are impermanent, whereas above it stationary spot and
stripe patterns can coexist over time. We have observed the striking
formation of spatial patterns during the evolution, but the isolated
ordered spot patterns don't emerge in the space.
\end{abstract}

%Use showkeys class option if keyword
\maketitle

\section{Introduction}
The dynamics of spontaneous spatial pattern formation, first
introduced to biology by Turing~\cite{AMT1952} five decades ago, has
recently been attracting attention in many subfields of biology to
describe various phenomena. Non-equilibrium labyrinthine patterns
are observed in chemical reaction-diffusion systems with a Turing
instability~\cite{OuyangQi} and in bistable reaction-diffusion
systems~\cite{PhysRevLett.72.2494,PhysRevE.51.1899}. Such dynamic
patterns in a two dimensional space have recently been introduced
into ecology~\cite{levin:45,AG1998,AG1999,HS2003,MB2004}. In the
past few years, geophysical patterns over a wide range of scales for
the vegetation have been presented and studied in the
Refs.~\cite{gilad:098105,PhysRevLett.87.198101,PhysRevE.66.010901,MR2002,HilleRisLambers}
using the Turing mechanisms.

In the epidemiology, one of the central goals of mathematical
epidemiology is to predict in populations how diseases transmit in
the space. For instance, the SARS epidemic spreads through 12
countries within a few weeks. The classical epidemic SIR model
describes the infection and recovery process in terms of three
ordinary differential equations for susceptibles (S), infected (I),
and recovered (R), which has been studied by many
researchers~\cite{WW2004,RMAnderson,2000Sci,Diekmann1991} and the
reference cited therein. These systems depend mainly on two
parameters, the infection rate and the recovery rate.

 A growing
body of work reports on the role of spatial patterns on evolutionary
processes in the host population
structure~\cite{WMarijn2004,MBoots1999,M.Boots02062004,CRJ2002,DARand1995,HY200,BOOTS2000}.
Recent studies have shown large-scale spatiotemporal patterns in
measles~\cite{BTGrenfell} and dengue fever
(DF)~\cite{Derek2004,vecchio:031913}. More dramatically the wave is
often caused by the diffusion (or invasion) of virus within the
populations in a given spatial region, thus generating periodic
infection, which has been observed in the occurrence of dengue
hemorrhagic fever (DHF) in Thailand~\cite{DATC2004}. Existing
theoretical work on pathogen evolution and spatial pattern formation
has focused on a model in which local invasion to the susceptible
hosts plays a central
role~\cite{MBoots1999,M.Boots02062004,BOOTS2000}. Projections of the
spatial spread of an epidemic and the interactions of human movement
at multiple levels with a response protocol will facilitate the
assessment of policy alternatives. Spatially-explicit models are
necessary to evaluate the efficacy of movement
controls~\cite{StevenRiley06202003,EUBANK2004}. A wide variety of
methods have been used for the study of spatially structured
epidemics, such as cellular
automata~\cite{liu:031110,Doran2005,fuks-2001-6},
networks~\cite{Bauch2003,PhysRevE.66.016128},
metapopulations~\cite{MJKeeling2000,Lloyd2004}, diffusion
equations~\cite{Caraco2002,PhysRevE.57.3622,Reluga2006}, and
integro-differential equations, which are useful tools in the study
of geographic epidemic spread. In particular, spatial models can be
used to estimate the formation of spatial patterns in large-scale
and the transmission velocity of diseases, and in turn guide policy
decisions.

This paper addresses how diffusive contacts and diffusive movement
affect the formation of spatial patterns in two dimensions. The
diffusion term is from the earlier work that tracing back to Fisher
and Kolmogorov. Noble applied diffusion theory to the spread of
bubonic plague in Europe~\cite{Noble1974}. Noble's model relies on
the assumptions that disease transmits through interactions between
dispersing individuals and infected individuals move in uncorrelated
random walks. In light of the Turing theoretical and study of recent
spatial models, we investigate the formation of spatial patterns in
the spatial SIR model based on the study of non-spatial SIR model
with constant removal rate of the infectives~\cite{WW2004}.

\section{Model}
\subsection{Basic model} We consider, as the basic model, the
following Susceptible-Infected-Recovery (SIR) model
\begin{subequations}\label{eq:2}
\begin{equation}\label{eq:2a}
\frac{dS}{dt}=A-dS-\lambda SI,
\end{equation}
\vspace{-0.5cm}
\begin{equation}\label{eq:2b}
  \frac{dI}{dt}=\lambda SI-(d+\gamma)I-h(I),
\end{equation}
\vspace{-0.5cm}
\begin{equation}\label{eq:2c}
\frac{dR}{dt}=\gamma I+h(I)-dR,
\end{equation}
\end{subequations}
where $S(t)$, $I(t)$, and $R(t)$ denote the numbers of susceptible,
infective, and recovered individuals at time $t$, respectively. $A$
is the recruitment rate of the population (such as growth rate of
average population size, a recover becomes an susceptible, immigrant
and so on), $d$ is the natural death rate of the population,
$\gamma$ is the natural recovery rate of the infective individuals,
and $\lambda$ is a measure of the transmission efficiency of the
disease from susceptibles to infectives. In Eq.~\eqref{eq:2}, $h(I)$
is the removal rate of infective individuals due to the treatment.
We suppose that the treated infectives become recovered when they
are treated in treatment sites. We also suppose that
\begin{equation}\label{eq:3}
h(I)=
\begin{cases} r, & {\rm for}\quad I>0,\\
0, & {\rm for}\quad I=0,
\end{cases}
\end{equation}
%
%   \makeatletter
%   \let\@@@alph\@alph
%   \def\@alph#1{\ifcase#1\or \or $'$\or $''$\fi}\makeatother
%   \begin{subnumcases}
%   {h(I)=}
%   r, & for $I>0$, \label{eq:3}\\
%   0, & for $I=0$,\nonumber\label{eq:a3}
%   \end{subnumcases}
%   \makeatletter\let\@alph\@@@alph\makeatother
where $r>0$ is constant and represents the capacity of treatment for
infectives. The detail about model~\eqref{eq:2} can be found in
Ref.~\cite{WW2004}.

\subsection{Spatial model}

Next we intend to add the spatial parts. Upto the first
approximation, the dispersal of individuals can be taken random, so
that Fick' law holds. This gives the flux terms as
\begin{equation}\label{eq:4}
    \frac{\partial S}{\partial t}=D_{ s}\nabla^2 S, \frac{\partial I}{\partial t}=D_{ i}\nabla^2
    I, \frac{\partial R}{\partial t}=D_{ r}\nabla^2 R,
\end{equation}
where $\nabla^2$ ($\nabla^2=\frac{\partial^2}{\partial
x^2}+\frac{\partial^2}{\partial y^2}$) is the Laplacian operator in
Cartesian coordinates. $D_{ s}$, $D_{ i}$, and $D_{ r}$ are the
diffusion coefficients of the susceptible, infective, and recovered,
respectively. Incorporating spatial terms into Eq.~\eqref{eq:2}, it
becomes
\begin{subequations}\label{eq:5}
\begin{equation}\label{eq:1a}
\frac{\partial S}{\partial t}=A-dS-\lambda SI+D_{ s}\nabla^2 S,
\end{equation}
\vspace{-0.5cm}
\begin{equation}\label{eq:1b}
  \frac{\partial I}{\partial t}=\lambda SI-(d+\gamma)I-h(I)+D_{ i}\nabla^2 I,
\end{equation}
\vspace{-0.5cm}
\begin{equation}\label{eq:1c}
\frac{ \partial R}{\partial t}=\gamma I+h(I)-dR+D_{ r}\nabla^2 R.
\end{equation}
\end{subequations}

Generally, we concern on the susceptible and infective individuals.
Moreover the Eqs.~\eqref{eq:1a} and \eqref{eq:1b} are independent of
the Eq.~\eqref{eq:1c} whose dynamic behavior is trivial when
$I(t_0)=0$ for some $t_0>0$. So it suffices to consider the
Eqs.~\eqref{eq:5a} and \eqref{eq:5b} with $I>0$. Thus, we restrict
our attention to the following reduced spatial model
\begin{subequations}\label{eq:5}
\begin{equation}\label{eq:5a}
\frac{\partial S}{\partial t}=A-dS-\lambda SI+D_{ s}\nabla^2 S,
\end{equation}
\vspace{-0.5cm}
\begin{equation}\label{eq:5b}
  \frac{ \partial I}{\partial t}=\lambda SI-(d+\gamma)I-r+D_{ i}\nabla^2
  I.
\end{equation}
\end{subequations}
It is assumed that all the parameters are positive constants from
the biological point of view.

\section{Theoretical analysis of spatial patterns and Results}

To study the mechanism of the formation of spatial patterns,
firstly, we analyze the stability criterion of the local system.
This can be obtained from the Ref.~\cite{WW2004}. The
system~\eqref{eq:5} has two positive equilibrium points if $R_0>1$
and $0<H<(\sqrt{R_0}-1)^2$, where $R_0=\frac{\lambda
A}{d(d+\gamma)}$ and $H=\frac{\lambda r}{d(d+\gamma)}$. The two
positive equilibrium points are $E_1=(S_1,I_1)$ and $E_2=(S_2,I_2)$,
where
\begin{eqnarray*}
  I_1 &=& \frac{d}{2\lambda}(R_0-1-H-\sqrt{(R_0-1-H)^2-4H}), \nonumber\\
   S_1&=& A/(d+\lambda I_1),\nonumber \\
  I_2 &=& \frac{d}{2\lambda}(R_0-1-H+\sqrt{(R_0-1-H)^2-4H}),\nonumber\\
  S_2&=& A/(d+\lambda I_2).  \nonumber
\end{eqnarray*}

Diffusion is often considered a stabilizing process, yet it is the
diffusion-induced instability in a homogenous steady state that
results in the formation of spatial patterns in a reaction-diffusion
system~\cite{AMT1952}. The stability of any system is expressed by
the eigenvalues of the system's Jacobian Matrix. The stability of
the homogenous steady state requires that the eigenvalues have
negative real parts. To ensure this negative sign, the trace of the
Jacobian matrix must be less than zero at steady state if the
determinant is greater than zero.

The Jacobian matrix of system~\eqref{eq:2} at $(S_2,I_2)$ is
\begin{eqnarray} \label{Jac}
J_2=\left(
\begin{array}{cc}
-d-\lambda I_2 & -\lambda S_2\\
\lambda I_2 & \lambda S_2-d-\gamma
\end{array}\right)\;.
\end{eqnarray}
From the Ref.~\cite{HQ2001}, we easily know that there are the
Turing space in the system~\eqref{eq:5} at point $E_2$, but at point
$E_1$ there is no Turing space.

\subsection{Stability of the positive equilibrium point in the spatial model}

In contrast to the local model, we employ the spatial model in a
two-dimensional (2D) domain, so that the steady-state solutions are
2D functions. Let us now discuss the stability of the positive
equilibrium point with respect to perturbations. Turing proves that
it is possible for a homogeneous attracting equilibrium to lose
stability due to the interaction of diffusion process. To check
under what conditions these Turing instabilities occur in the
model~\eqref{eq:5}, we test how perturbation of a homogeneous
steady-state solution behaves in the long-term limit. Here we choose
perturbation functions consisting of the following 2D Fourier modes

\addtocounter{equation}{1}
\begin{align}\label{eq:7a}
\hat{s}={\rm exp}((k_x x+k_y y)i+\delta_k t), \tag{\theequation a}\\
\hat{i}={\rm exp}((k_x x+k_y y)i+\delta_k t). \tag{\theequation
b}\label{eq:7b}
\end{align}

Since we will work with the linearized form of Eq.~\eqref{eq:5} and
the Fourier modes are orthogonal, it is sufficient to analyze the
long-term behavior of an arbitrary Fourier mode.

After substituting $S=S_2+\hat{s}$ and $I=I_2+\hat{i}$ in
Eq.~\eqref{eq:5}, we linearize the diffusion terms of the equations
via a Taylor-expansion about the positive equilibrium point
$E_{2}(S_2,I_2)$ and obtain the characteristic equation
\begin{eqnarray} \label{Jac1}
(J_{\rm sp}-\delta_k \textbf{I})\cdot\left(
\begin{array}{c}
\hat{s} \\
\hat{i}
\end{array}\right)=0\;,
\end{eqnarray}
with
\begin{eqnarray} \label{Jac2}
J_{\rm sp}=\left(
\begin{array}{cc}
 j_{11}-D_{ s}k^2& j_{12}\\
 j_{21}& j_{22}-D_{ i}k^2
\end{array}\right)\;,
\end{eqnarray}
here $j_{11}=-d-\lambda I_2$, $j_{12}=-\lambda S_2$, $j_{21}=\lambda
I_2$, and $j_{22}=\lambda S_2-d-\gamma$. $k^2=k_{x}^2+k_{y}^2$ and
$k$ represents the wave numbers.

To find Turing instabilities we must focus on the stability
properties of the attracting positive equilibrium point
$E_{2}(S_2,I_2)$. The loss of stability occurs if at least one of
the eigenvalues of the matrix $J_{\rm sp}-\delta_k \textbf{I}$
crosses the imaginary axis. From the Eqs.~\eqref{Jac1} and
\eqref{Jac2}, we can obtain the characteristic equation
\begin{equation}\label{eq:10}
    {\rm det}(J_{\rm sp}-\delta_k \textbf{I})=\delta_{k}^2-{\rm
    tr}({J_{\rm sp}})\delta_k +{\rm det}(J_{\rm sp})=0,
\end{equation}
where ${\rm tr}({J_{\rm sp}})={\rm tr}(J_2)-(D_{ s}+D_{ i})k^2$ and
${\rm det}(J_{\rm sp})={\rm det}(J_2)-k^2(j_{11}D_{ i}+j_{22}D_{
s})+k^4D_{ s}D_{ i}$. Taking  ${\rm tr}(J_2)>{\rm tr}({J_{\rm sp}})$
into account, we can obtain that for saddles and attractors (both
with respect to the non-spatial model) a change of stability
coincides with a change of the sign of ${\rm det}(J_{\rm sp})$.

\begin{figure*}
\scalebox{0.55}[0.55]{\includegraphics{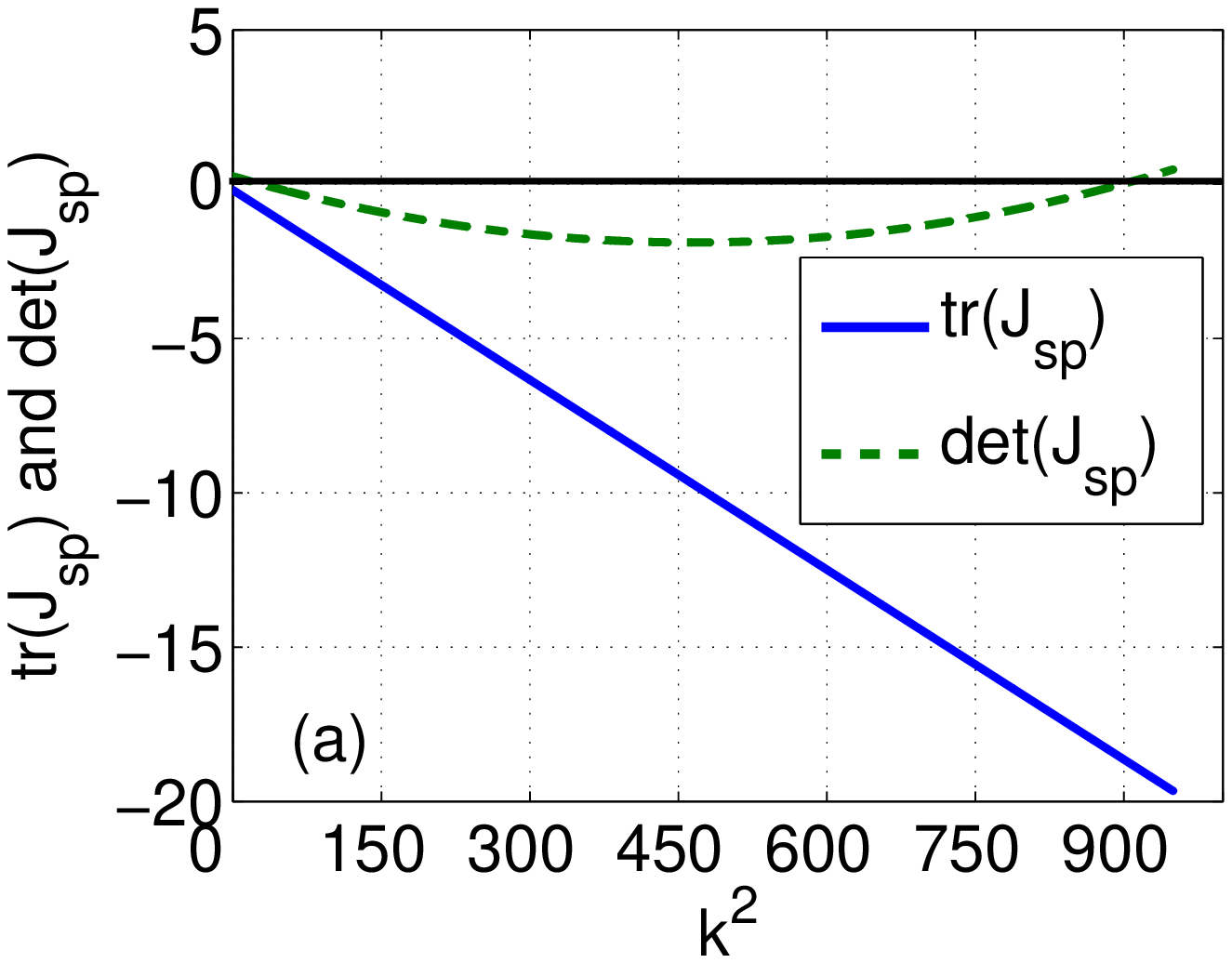}}% Here is how to import EPS art
\scalebox{0.55}[0.55]{\includegraphics{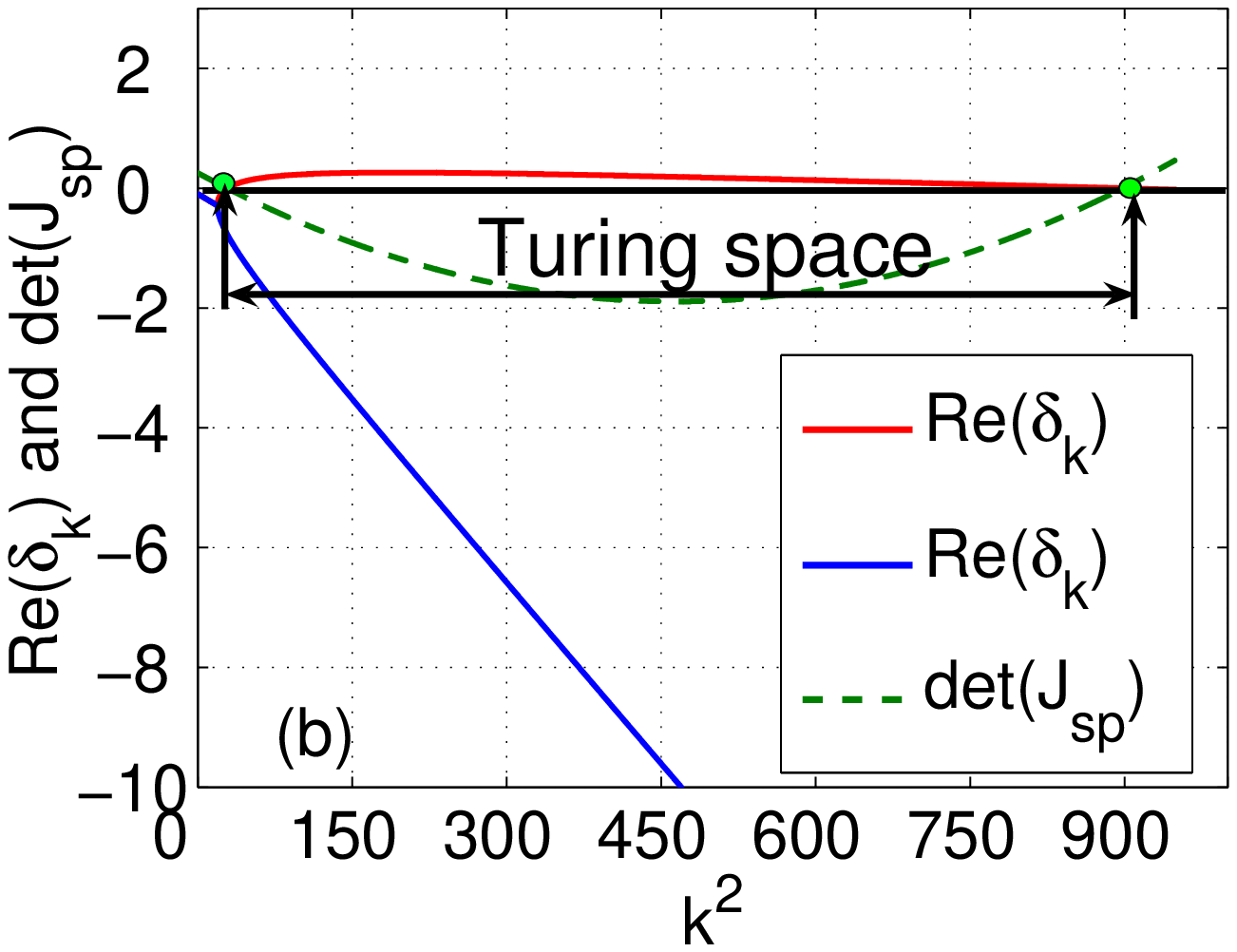}}% Here is how to import EPS art
\caption{\label{fig1}(Color online) This graphes illustrate the
eigenvalues of the spatial model~\eqref{eq:5} at positive
equilibrium point $E_{2}(S_2,I_2)$, and the loss of stability occurs
relation to the limit range wave numbers. (a) For diffusion-driven
instability arise both ${\rm tr}(J_{\rm sp})$ and ${\rm det}(J_{\rm
sp})$ must be negative for some range of $k^2$; (b) The real parts
of the eigenvalues of matrix~\eqref{Jac2} at positive equilibrium
$(S_2,I_2)$. Model parameters used here are: $A=3$, $d=0.3$,
$\lambda=0.35$, $r=0.5$, $\gamma=0.8$, $D_{ s}=0.02$, and $D_{
i}=0.0005$. }
\end{figure*}

Doing some calculations we find that a change of the sign of ${\rm
det}(J_{\rm sp})$ occurs when $k^2$ takes the critical values
\addtocounter{equation}{1}
\begin{align}\label{eq:11a}
k^2_{-}=\frac{j_{11}D_{ i}+j_{22}D_{ s}-\sqrt{(j_{11}D_{
i}+j_{22}D_{ s})^2-4D_{ s}D_{ i}{\rm
det}(J_2)}}{2D_{ s}D_{ i}} \tag{\theequation a},\\
k^2_{+}=\frac{j_{11}D_{ i}+j_{22}D_{ s}+\sqrt{(j_{11}D_{
i}+j_{22}D_{ s})^2-4D_{ s}D_{ i}{\rm det}(J_2)}}{2D_{ s}D_{ i}}.
\tag{\theequation b}\label{eq:11b}
\end{align}
In particular, we have
\begin{equation}\label{eq:12}
 {\rm det}(J_{\rm sp})<0 \Longleftrightarrow  k_{-}^2<k^2<k_{+}^2.
\end{equation}
If both $k_{-}^2$ and $k_{+}^2$ exist and have positive values, they
limit the range of instability for a local stable equilibrium. We
refer to this range as the Turing Space (or Turing Region, see
Fig.~\ref{fig1}).

In Fig.~\ref{fig1}, the real parts of the eigenvalues of the spatial
model~\eqref{eq:5} at positive equilibrium point $E_{2}(S_2,I_2)$
are plotted. From the Eqs.~\eqref{eq:7a} and \eqref{eq:7b}, we know
that the parameter $\delta_k$ can either be a real number or a
complex number. If it is a real number, the spatial patterns will
emerge and be stable over time and otherwise the spatial patterns
will die away quickly. In both case, the sign of the real parts of
$\delta_k$ (written as $Re(\delta_k)$) is crucially important to
determine whether the patterns will grow or not. In particular if
$Re(\delta_k)>0$, spatial patterns will grow in the linearized
system because $|e^{\delta_k}|>1$, but if $Re(\delta_k)<0$ the
perturbation decays because $|e^{\delta_k}|<1$ and the system
returns to the homogeneous steady state. Further details concerning
linear stability analysis can be found in Ref.~\cite{JDM1993}. The
Fig.~\ref{fig1} presents the typical situation of Turing
instability. With respect to homogenous perturbations,
$E_{2}(S_2,I_2)$ is stable at first, but when $k^2$ increases, one
eigenvalue changes its sign (when $k^2$ arrives at $k_{-}^2$.), the
instability occurs. The instability exist until $k^2$ reaches
$k_{+}^2$. When $k^2$ is over $k_{+}^2$, $(S_2,I_2)$ returns
stability again. Thus the Turing space is bounded between $k_{-}^2$
and $k_{+}^2$.

\begin{figure*}
\scalebox{0.55}{\includegraphics{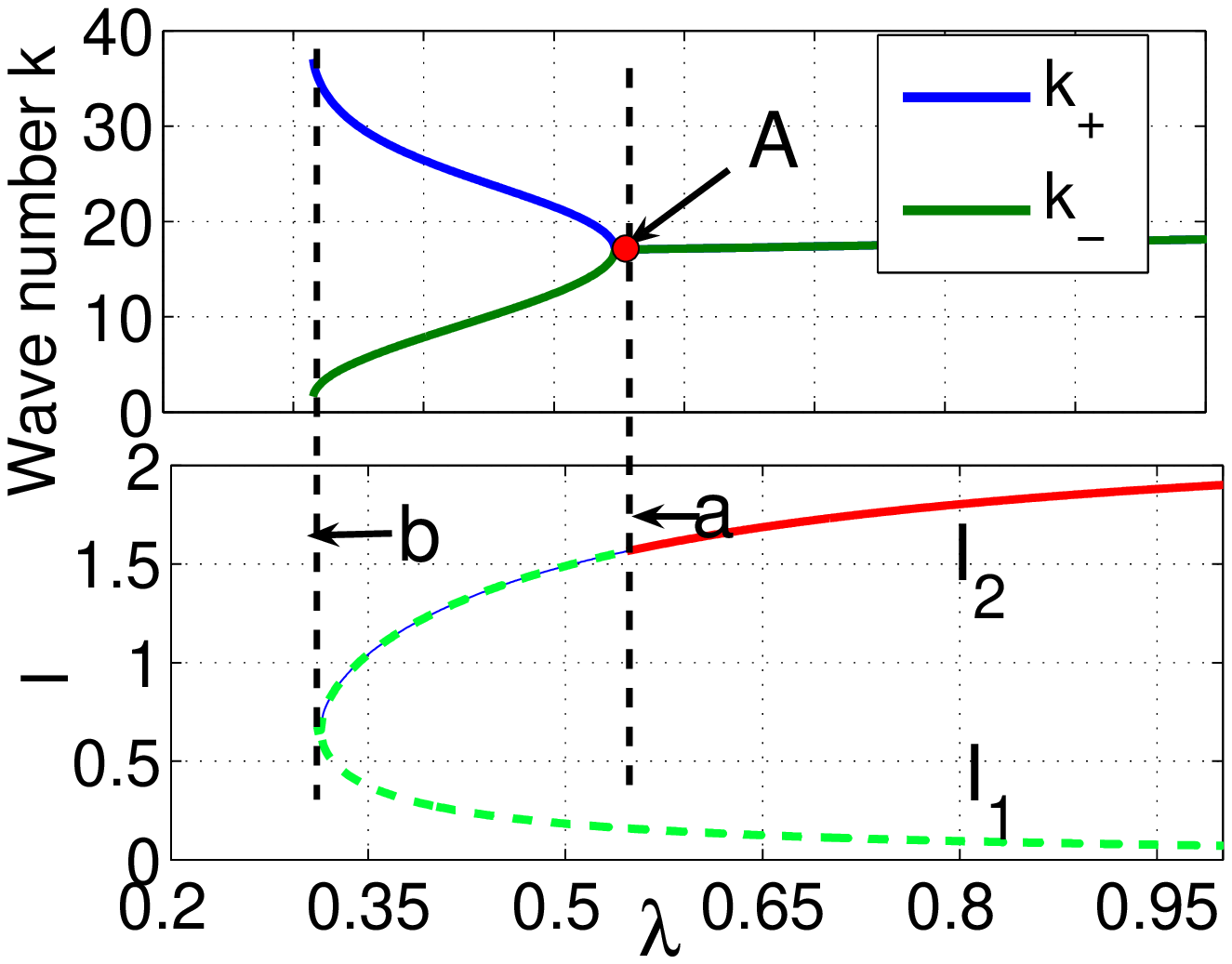}}% Here is how to import EPS art
\scalebox{0.55}{\includegraphics{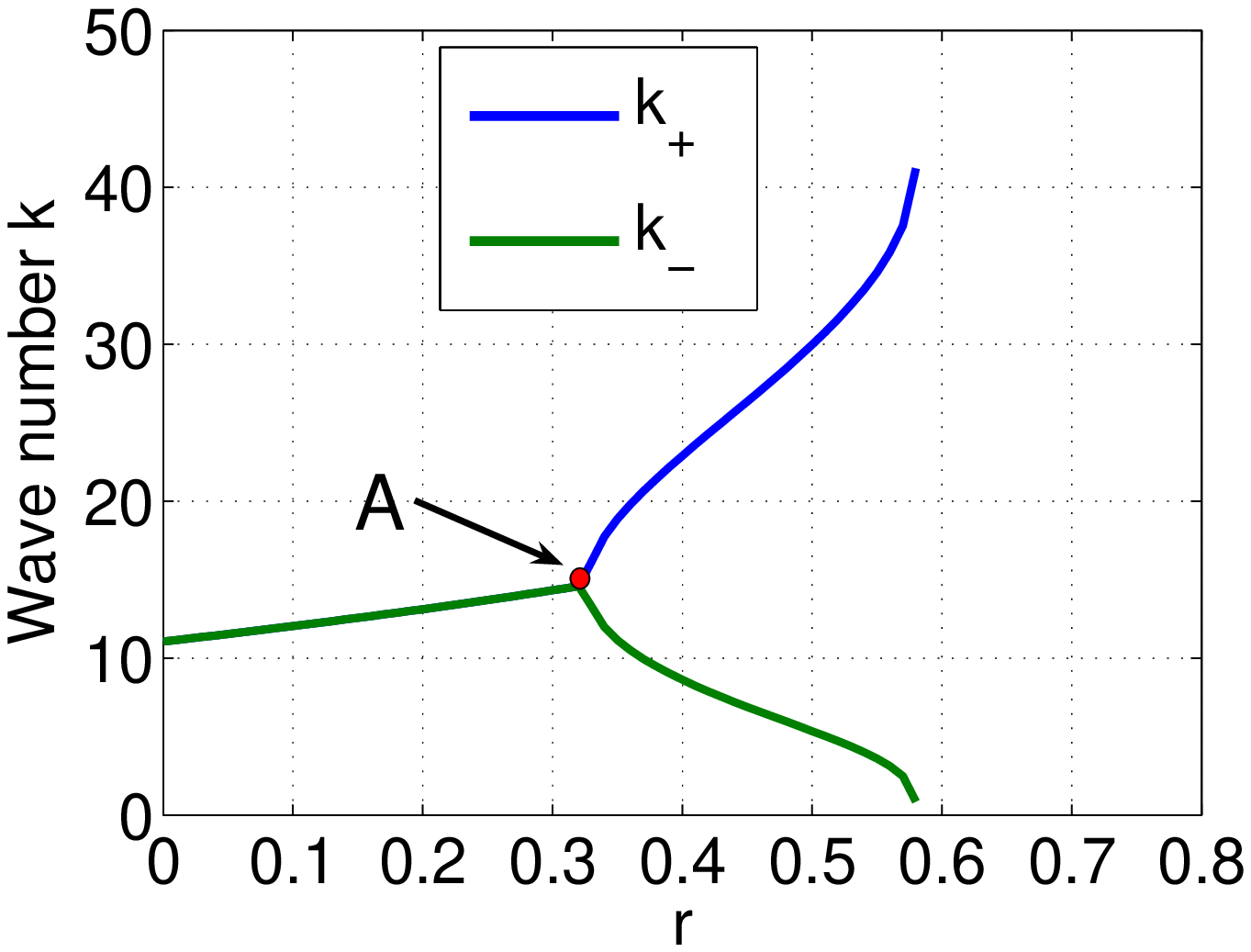}}% Here is how to import EPS art
\caption{\label{fig2}(Color online) The graphes illustrate Turing
space versus the parameter $\lambda$ and $r$, respectively. In the
Turing space $k_-$ and $k_+$ curves are shown respectively. (left)
The Bifurcation diagrams show the formation of the two stationary
solutions of Eq.~\eqref{eq:5} with fixed $r$ and varying $\lambda$,
and values of the parameters are $A=3$, $d=0.3$, $r=0.5$,
$\gamma=0.8$, $D_{ s}=0.02$, and $D_{ i}=0.0005$; In the lower
diagram, the green dashed line represents the loss of stability for
the positive equilibrium. The red line represents the stable
equilibrium. (right) The values of the parameters are $A=3$,
$d=0.3$, $\lambda=0.35$, $\gamma=0.8$, $D_{ s}=0.02$, and $D_{
i}=0.0005$.}
\end{figure*}

The change of the bounds $k_{-}$ and $k_{+}$ with respect to the
variation of the $\lambda$ and $r$ are illustrated in the
Fig.~\ref{fig2}, respectively. The typical feature of Turing space
in the model~\eqref{eq:5} can be observed in Fig.~\ref{fig2}. The
Turing space is limited by two different bounds. From
Fig.~\ref{fig2}(left-top) we can see that $k_-$ and $k_+$ converge
in one point A which corresponds to the critical value, $\lambda_c$.
Beyond right bound (line \emph{a}), the $E_{2}(S_2,I_2)$ exists and
is stable. The left bound (line \emph{b}) of the Turing space shows
an ``open end", which corresponds to the saddle-node in the
bifurcation plot (see Fig.~\ref{fig2}(left-bottom)) for the
model~\eqref{eq:2} and the equilibrium point $E_{2}(S_2,I_2)$ does
not exist under this bound. This figure shows the solutions of
equilibrium points $I_{1}$ and $I_{2}$, where the solid curves
represent attractors, dashed curve represents the repellers and
saddles, the dotted line \emph{a} represents the periodic points.
This diagram explains the $E_{2}(S_2,I_2)$ changes from repeller to
attractor, and an unstable orbit of periodical points emerges. From
the Fig.~\ref{fig2}(right), the curves indicates that $k_-$ and
$k_+$ converge in one point (A). Below that bound, the
$E_{2}(S_2,I_2)$ exists and is stable. The right bound of the Turing
space also shows an ``open end".

\begin{figure}
\scalebox{0.55}[0.55]{\includegraphics{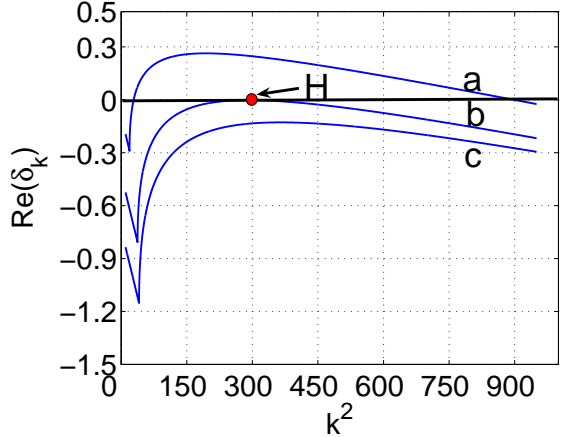}}% Here is how to import EPS art
%\scalebox{0.30}[0.30]{\includegraphics{picture/fig2b.eps}}% Here is how to import EPS art
\caption{\label{fig5}(Color online) Basic dispersion relation giving
the growth rate $Re(\delta_{k})$ as a function of the wavenumber
$k^2$. The mode become marginal at the point (H,
$\lambda=\lambda_c$) a finite-$k^2$ (Turing) mode. The parameter
values are $A=3$, $d=0.3$, $r=0.5$, $\gamma=0.8$, $D_{ s}=0.02$, and
$D_{ i}=0.0005$.}
\end{figure}

Comparing to the two graphes in Fig.~\ref{fig2}, one can obtain that
the parameters $\lambda$ and $r$ have a similar dynamical behavior
in the system for the Turing-bifurcation, but their effects are
opposite. We use the parameter $\lambda$ as the Turing-bifurcation
parameter in present paper. Fig.~\ref{fig5} shows growth rate curves
of the spatial patterns, where at bifurcation (curve \emph{b}),
$\lambda=\lambda_{c}\approx 0.547$ (The threshold $\lambda_{c}$ can
be derived analytically, see the Appendix. The analytical and
numerical values of $\lambda_c$ are approximately equal.), from
spatially uniform to spatially heterogenous the critical wave number
(point H) is $k_c=\sqrt{k_{xc}^2+k_{yc}^2}$. The curves \emph{a} and
\emph{c} correspond to the parameter $\lambda=0.35$  below the
$\lambda_c$, and $\lambda=0.8$  above the $\lambda_c$, respectively.
The spatial patterns are generated when $\lambda$ passes through the
critical Turing-bifurcation point $\lambda_c$. And for
$\lambda<\lambda_c$ there is a finite range of unstable wave numbers
which grow exponentially with time, $O({\rm exp}(\delta_{k}t))$,
where $\delta_{k}>0$ for a finite range of $k$.

The stable characteristics of $E_{2}(S_2,I_2)$ can be changed by the
variation of parameter $\lambda$: A sufficiently high increase of
$\lambda$ will turns $E_{2}(S_2,I_2)$ into an attractor. When
changing its characteristics, $E_{2}(S_2,I_2)$ traverses a
subcritical Hopf bifurcation and an unstable periodic orbit emerges
(the dotted line \emph{a} in Fig.~\ref{fig2}(left)). Surprisingly,
the latter is not necessarily true, if effect of diffusion comes to
play.

\subsection{Spatial patterns of the spatial model}

The numerical simulations are performed in this section for the
spatial model~\eqref{eq:5} in two dimensions. During the simulation,
the periodic boundary conditions are used and part of the parameter
values can be determined following Ref.~\cite{WW2004} (see the
Fig.~\ref{fig1} and \ref{fig2}). We assume that the homogeneous
$E_{2}(S_2,I_2)$ distributions are in uniform states for each start
of the simulation. To induce the dynamics that may lead to the
formation of spatial patterns, we perturb the $I$-distribution by
small random values.

\begin{figure}[htp]
\scalebox{0.30}[0.30]{\includegraphics[angle=90]{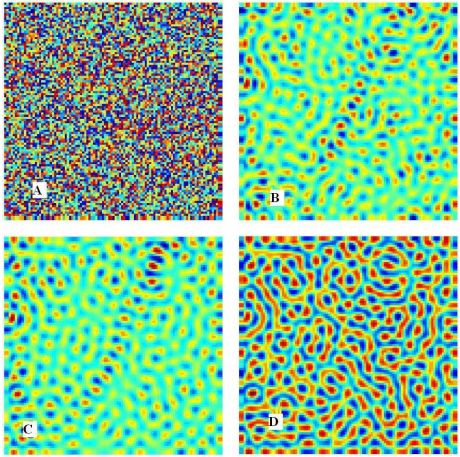}}% Here is how to import EPS art
%\scalebox{0.30}[0.30]{\includegraphics{picture/fig3b.eps}}% Here is how to import EPS art
%\scalebox{0.30}[0.30]{\includegraphics{picture/fig3c.eps}}
\caption{\label{fig3}(Color online) Snapshots of contour pictures of
the time evolution of $I(x,y,t)$ at different instants. (a)-(c)
Numerical results in $100\times100$ sites. The parameter values are
$A=3$, $d=0.3$, $r=0.5$, $\gamma=0.8$, $\lambda=0.65$, $D_{
s}=0.02$, $D_{ i}=0.0005$, and $\Delta x=\Delta y=0.05$. (A) 0
iteration; (B) 5000 iterations; (C) 30000 iterations; (D) 40000
iterations. [Additional movie formats available from the author]}
\end{figure}

\begin{figure}[htp]
\scalebox{0.30}[0.30]{\includegraphics[angle=90]{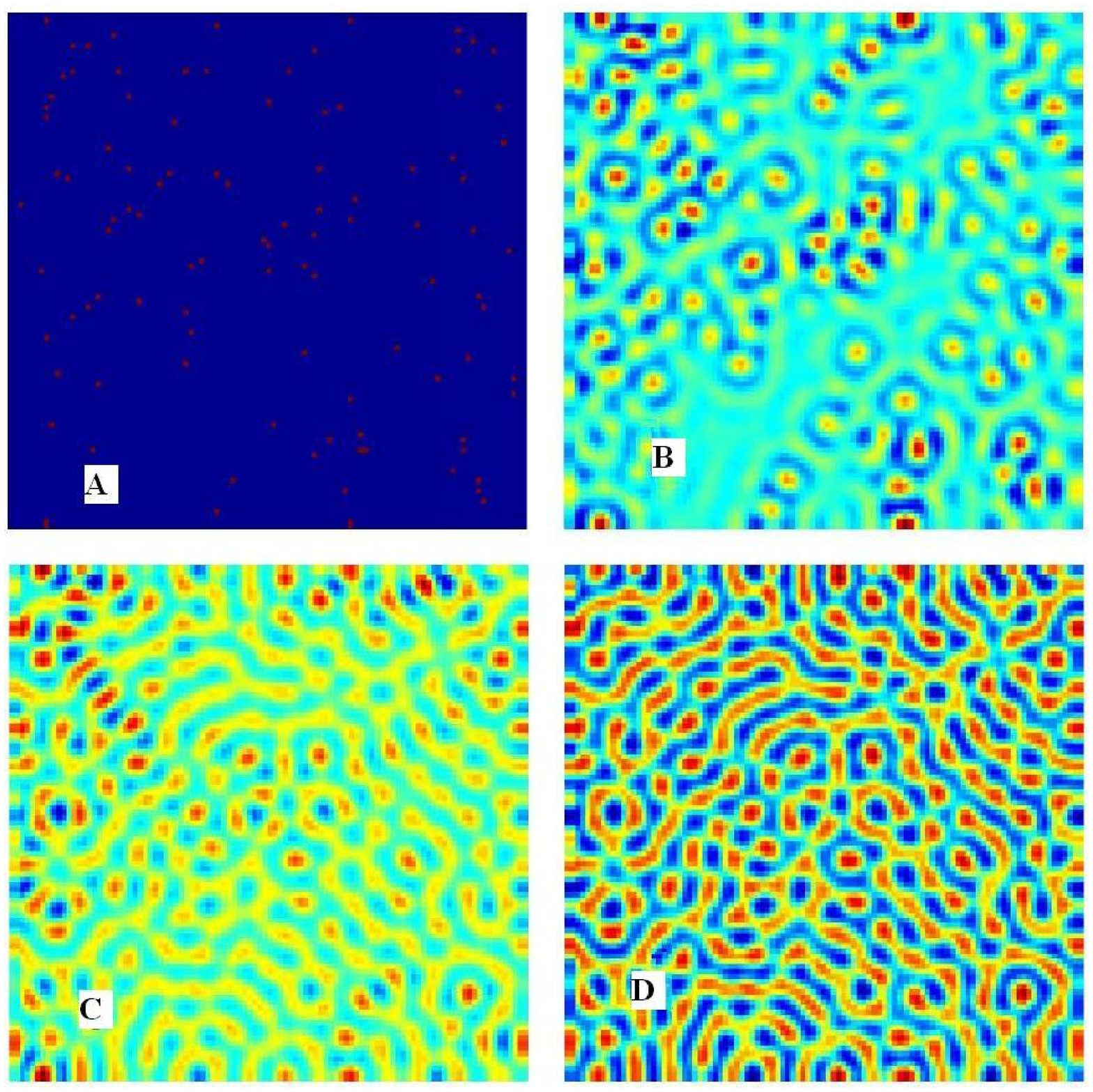}}% Here is how to import EPS art
%\scalebox{0.30}[0.30]{\includegraphics{picture/fig4b.eps}}% Here is how to import EPS art
%\scalebox{0.30}[0.30]{\includegraphics{picture/fig4c.eps}}
\caption{\label{fig4} (Color online) Snapshots of contour pictures
of the time evolution of $I(x,y,t)$ at different instants. The
parameters values are the same as Fig.~\ref{fig3}. (A) 0 iteration;
(B) 5000 iterations; (C) 40000 iterations; (D) 42000 iterations.
[Additional movie formats available from the author]}
\end{figure}

\begin{figure}[htp]
\scalebox{0.30}[0.30]{\includegraphics[angle=90]{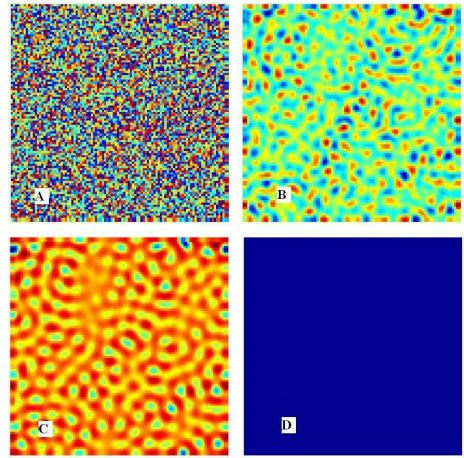}}% Here is how to import EPS art
%\scalebox{0.30}[0.30]{\includegraphics{picture/fig4b.eps}}% Here is how to import EPS art
%\scalebox{0.30}[0.30]{\includegraphics{picture/fig4c.eps}}
\caption{\label{fig6} (Color online) Snapshots of contour pictures
of the time evolution of $I(x,y,t)$ at different instants. The
parameters values are the same as Fig.~\ref{fig3} but $\lambda=0.5$.
(A) 0 iteration; (B) 1500 iterations; (C) 8500 iterations; (D) 9400
iterations. [Additional movie formats available from the author]}
\end{figure}

We study the spatial model~\eqref{eq:5} by performing stable
analysis of the uniform solutions and by integrating
Eqs.~\eqref{eq:5a} and \eqref{eq:5b} numerically at different values
of $\lambda$ on a grid of $100 \times 100$ sites by a simple Euler
method with a time step of $\Delta t=0.01$. The results for the
infected are summarized below in two dimensions. The model has a
uniform free-disease state (no infected) for all constant values of
$\lambda$, represented by the solution $S=A/d, I=0$. The
free-disease state is stable when $\lambda<\lambda_{c'}$. Here
$\lambda_{c'}$ is a critical value (or threshold ) corresponded by
the dotted line \emph{b} in Fig.~\ref{fig2} (left). Above
$\lambda_{c'}$ two new states appear, shown in Fig.~\ref{fig2}
(left) as line $I_1$ and $I_2$. The state $I_2$ represents a
uniformly distributed population with infected density monotonically
increasing with $\lambda$. It is instable only for relative values
of $\lambda$, $\lambda_{c'}<\lambda<\lambda_c$ and regains stability
when $\lambda>\lambda_c$, where the infected density is high. The
types of spatial patterns are depended on the range of parameter
$\lambda$ as in Figs.~\ref{fig3}, \ref{fig4}, and \ref{fig6}. We
have made movies of Figs.~\ref{fig3}, ~\ref{fig4} and ~\ref{fig6},
respectively, as supplementary materials.

We test several different initial states within the linear regime
and the nonlinear regime respectively. Figs.~\ref{fig3} and
\ref{fig4} show that stationary stripe and spot patterns emerge
mixtedly in the distribution of the infected population density,
where the $\lambda$ is above $\lambda_c$ in the linear regime. The
initial state of Fig.~\ref{fig3} is the random perturbation of the
stable uniform infected state. The initial state of Fig.~\ref{fig4}
consists of a few spots (100 scattered spots). Values of the
parameters are the same in both two figures.  In the linear regime,
the result shows that the stripes and spots which describe
asymptotic patterns for the spatial model~\eqref{eq:5} converge at
long times. Different initial states may lead to the same type of
asymptotic patterns, but the transient behaviors are obviously
different (compare the Fig.~\ref{fig3}(b) with Fig.~\ref{fig4}(b)).
Unfortunately, the linear predictions are not accurate in the
nonlinear regime.

Fig.~\ref{fig6} shows the snapshots of spatial patterns when the
$\lambda$ is between $\lambda_{c'}$ and $\lambda_c$. One can see
that the spatial patterns resemble in Fig.~\ref{fig3} and
Fig.~\ref{fig6} at the beginning phase, but differ in the middle and
last phases. In Fig.~\ref{fig6}, the spatial patterns appear
spotted, holed and labyrinthine states in the middle phases, and the
spatial patterns appear uniform spatial states in the last phase.

\begin{figure}
\scalebox{0.3}[0.3]{\includegraphics[angle=90]{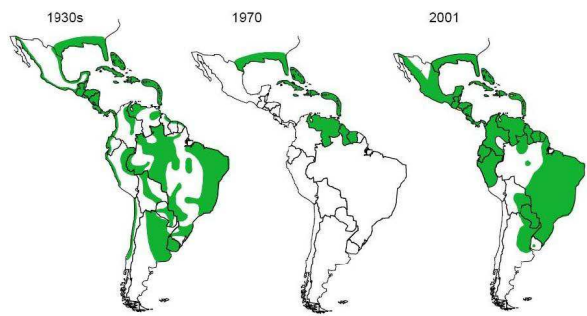}}% Here is how to import EPS art
%\hfill\scalebox{0.2}[0.2]{\includegraphics{picture/fig8.eps}}
%\scalebox{0.30}[0.30]{\includegraphics{picture/fig2b.eps}}% Here is how to import EPS art
\caption{\label{fig7}(Color online) Field observations of DHF
spatial patters. Reprinted form Trends In Microbiology,
Ref.~\cite{Gubler}, Copyright (2002), with permission from Elsevier.
\emph{Aedes aegypti} distribution in the Americas: 1930s, 1970 and
2001.}
\end{figure}

To explain spatial patterns arising from the spatial model, here we
present some observations of the spatial and temporal dynamics of
dengue hemorrhagic fever epidemics. Dengue fever (DF) is an old
disease that became distributed worldwide in the tropics during the
18th and 19th centuries. Fig.\ref{fig7} shows spotted and
labyrinth-like spatial patterns of DHF from the field
observations~\cite{Gubler}. By comparing the Figs.~\ref{fig3} and
\ref{fig4} with Fig.~\ref{fig7}, our results simply capture some key
features of the complex variation and explain the observation in
spatial structure to most vertebrate species, including humans. In
Figs.~\ref{fig3} and \ref{fig4}, the steady spatial patterns
indicate the persistence of the epidemic in the space. This result
well agrees with the field observation. More examples of the spatial
patterns of epidemic, such as HIV, poliovirus, one can find in
Refs.~\cite{Williams,NicholasC11172006}. In the light of recent work
of emphasizing the existence of `small world' networks in human
population, our results are also consistent with M. Boots and A.
Sasaki's conclusion that if the world is getting `small'--as
populations become more connected--disease may evolve higher
virulence~\cite{MBoots1999}.

\section{ Discussion}

In our paper, the labyrinthine patterns are found in the spatial
epidemic model driven by the diffusion. The spatial epidemic model
comes from the classical non-spatial SIR model which assumes that
the epidemic time scale is so tiny related to the demographic time
scale that demographic effects may be ignored. But here in the
spatial model we take the births and deaths into account. The
spatial diffusive epidemic model is more realistic than the
classical model. For instance, the history of bubonic plague
describes the movement of the disease from place to place carried by
rats. The course of an infection usually cannot be modeled
accurately without some attention to its spatial spread. To model
this would require partial differential equations (PDE), possibly
leading to descriptions of population waves analogous to disease
waves which have often been observed. Here our spatial model is
established from a basic dynamical `landscape' rather than other
perturbations, including environmental stochastic variations. From
the analysis of the Turing space and numerical simulations one can
see that the attracting positive equilibrium will occur instability
driven by the diffusion and the instability leads to the
labyrinthine patterns within the Turing space. This may explain the
prevalence of disease in large-scale geophysics. The positive
equilibrium is stable in the non-spatial models, but it may lose its
stability with respect to perturbations of certain wave numbers and
converge to heterogeneous distributions of populations. It is
interesting that we have not observed the isolated spots patterns in
the spatial epidemic model~\eqref{eq:5}.

The model~\eqref{eq:5} is introduced in a general form so that it
has broad applications to a range of interacting populations. For
example, it can be applied to diseases such as measles, AIDS, flu,
etc. Our paper focuses on the deterministic reaction-diffusion
equations. However, recent study shows noise plays an important role
on the epidemic model~\cite{RKuske,JonathanDushoff11302004}, which
indicate that the noise induces sustained oscillations and coherence
resonance in the SIR model.

\begin{acknowledgments}
This work was supported by the National Natural Science Foundation
of China under Grant No. 10471040 and the Natural Science Foundation
of Shan'xi Province Grant No. 2006011009. We are grateful for the
meaningful suggestions of the two anonymous referees.
\end{acknowledgments}

\appendix
\section{Appendixes}

Considering the dispersive relation associated with
Eqs.~\eqref{eq:7a} and \eqref{eq:7b},
%\begin{align}\label{eq:app1a}
%\hat{s}={\rm exp}((k_x x+k_y y)i+\delta_k t), \tag{\theequation a}\\
%\hat{i}={\rm exp}((k_x x+k_y y)i+\delta_k t), \tag{\theequation
%b}\label{eq:app1b}
%\end{align}
the functions of $\delta_k$ for the spatial model are defined by the
characteristic Eq.~\eqref{eq:10}. Now $Re(\delta_k)$ predict the
unstable wave modes from Eq.~\eqref{eq:10}. One can estimate the
most unstable wave number and the critical value of the bifurcation
parameter by noticing that at the onset of the instability
$\delta_{k}(k_c)=0$. Thus the constant term in Eq.~\eqref{eq:10}
must be zero at $k_c$. In the case of the spatial model this
condition is a second order equation on $k_{c}^2$, i.e.,
\begin{eqnarray}\label{eq:app2}
   && D_s D_i k_{c}^4-k_{c}^2((-d-\lambda I_2)D_i+(\lambda S_2-d-\gamma)D_s) \nonumber\\
   && +(-d-\lambda I_2)(\lambda S_2
    -d-\gamma)+\lambda^2 S_2 I_2=0.
\end{eqnarray}
And as a result the most unstable wave number is given by
$\frac{(-d-\lambda I_2)D_i+(\lambda S_2-d-\gamma)D_s}{2D_s D_i}$.
The critical Turing-bifurcation parameter value, which corresponds
to the onset of the instability is defined by Eq.~\eqref{eq:app2}.
In the spatial model $\lambda$ is the bifurcation parameter
adjusting the distance to the onset of the instability. The
discriminant of Eq.~\eqref{eq:app2} equals zero for $\lambda_c$ and
an instability takes place for $\lambda<\lambda_c$. Then, we have
\begin{eqnarray}\label{eq:app3}
   \lambda_c&=&\frac{(D_s S_2+D_i I_2+2 \sqrt{D_s D_i S_2 I_2})(D_s\gamma+D_s d-D_i d)}
   {(D_i I_2-D_s S_2)^2}, \nonumber\\
%   \text{or}\\
%   \lambda_{c'}&=&\frac{(D_s S_2+D_i I_2-2 \sqrt{D_s D_i S_2 I_2})(D_s\gamma+D_s d-D_i d)}
%   {(D_i I_2-D_s S_2)^2}, \nonumber\\
\end{eqnarray}
where $S_2=S_2(\lambda_c)$ and $I_2=I_2(\lambda_c)$. We can
calculate $\lambda_c$ from the Eq.~\eqref{eq:app3} by the computer.

%\newpage %Just because of unusual number of tables stacked at end
%\bibliography{apssamp1}% Produces the bibliography via BibTeX.

\end{document}